\documentclass[preprint,prb]{revtex4-1}

\usepackage{bm}
\usepackage{graphicx}
\usepackage{color}

\begin{document}

\title{U(1) and SU(2) quantum dissipative systems: The Caldeira-Leggett vs. the Amegaokar-Eckern-Sch\"on approaches}

\author{Alexander \surname{Shnirman}$^{1,2}$, Arijit \surname{Saha}$^3$, Igor S. \surname{Burmistrov}$^{2,4}$, Mikhail N.  \surname{Kiselev}$^5$, Alexander \surname{Altland}$^6$, Yuval \surname{Gefen}$^{7,8}$}
\affiliation{$^1$Institut f\"ur Theorie der Kondensierten Materie, Karlsruhe Institute of Technology, D-76128 Karlsruhe, Germany}
\affiliation{$^2$L.D. Landau Institute for Theoretical Physics RAS, Kosygina street 2, 119334 Moscow, Russia}
\affiliation{$^3$Institute of Physics, Sachivalaya Marg, Bhubaneswar, Orissa, 751005, India}
\affiliation{$^4$Moscow Institute of Physics and Technology, 141700 Moscow, Russia} 
\affiliation{$^5$International Center for Theoretical Physics, Strada Costiera 11, I-34014 Trieste, Italy}
\affiliation{$^6$Institut f\"ur Theoretische Physik, Universit\"at zu K\"oln, Z\"ulpicher Str. 77, D-50937 K\"oln, Germany}
\affiliation{$^7$Department of Condensed Matter Physics, Weizmann Institute of Science, 76100 Rehovot, Israel}
\affiliation{$^8$Institut f\"ur Nanotechnologie, Karlsruhe Institute of Technology, 76021 Karlsruhe, Germany}

\date{\today}

\begin{abstract}
There are two paradigmatic frameworks for treating quantum systems coupled to a dissipative environment: the Caldeira-Leggett 
and the Ambegaokar-Eckern-Sch\"on approaches. Here we recall the differences between them, and explain the consequences when each is applied to a zero dimensional spin (possessing an SU(2) symmetry) in a dissipative environment (a dissipative quantum dot near or beyond the Stoner instability point).
\end{abstract}

\maketitle

\vskip 1cm

The diagrammatic technique for non-equilibrium systems developed in the
pioneering works of Schwinger and Keldysh plays a predominant role in
theoretical condensed matter physics~\cite{Schwinger61,Keldysh65}. It is designed to tackle real time 
evolution of systems at and away from equilibrium. Following the developments of the last two
decades~\cite{KamenevAndreev,KamenevBook,AltlandBook}, it now provides a non-perturbative tool to tackle interaction induced strong correlations in quantum many-body systems. In this paper we discuss an important prototypical problem, a quantum zero dimensional degree of freedom in a dissipative environment, in which the Keldysh technique is of tremendous use, providing insight into the physics involved. 

\section{General perspective}
We consider here the dynamics of a quantum system  coupled to a dissipative environment. The resulting equation-of-motion is stochastic, which can be formulated on any of the following three levels: (i) a fully classical Langevin equation, where both the variables are classical (expectation values of observables) and the frequency range of interest is $\hbar \omega< k_B T$.  For  Ohmic dissipation the noise spectrum is white; (ii) a semi-classical hybrid description, within which the variables are still classical coordinates, but one acknowledges the fact that  the noise may 
be quantum, having high frequency component, $\hbar \omega> k_B T$ (Ref.~\cite{ASchmid82_Langevin}); (iii) a full-fledged quantum mechanical description, according to which the noise may contain high frequency quantum components, and the variables of the quantum Langevin equation are operators within the Heisenberg description. This approach is practiced, say, in the field of quantum optics~\cite{GardinerZollerBook}.

\begin{figure}
\includegraphics[scale=.5]{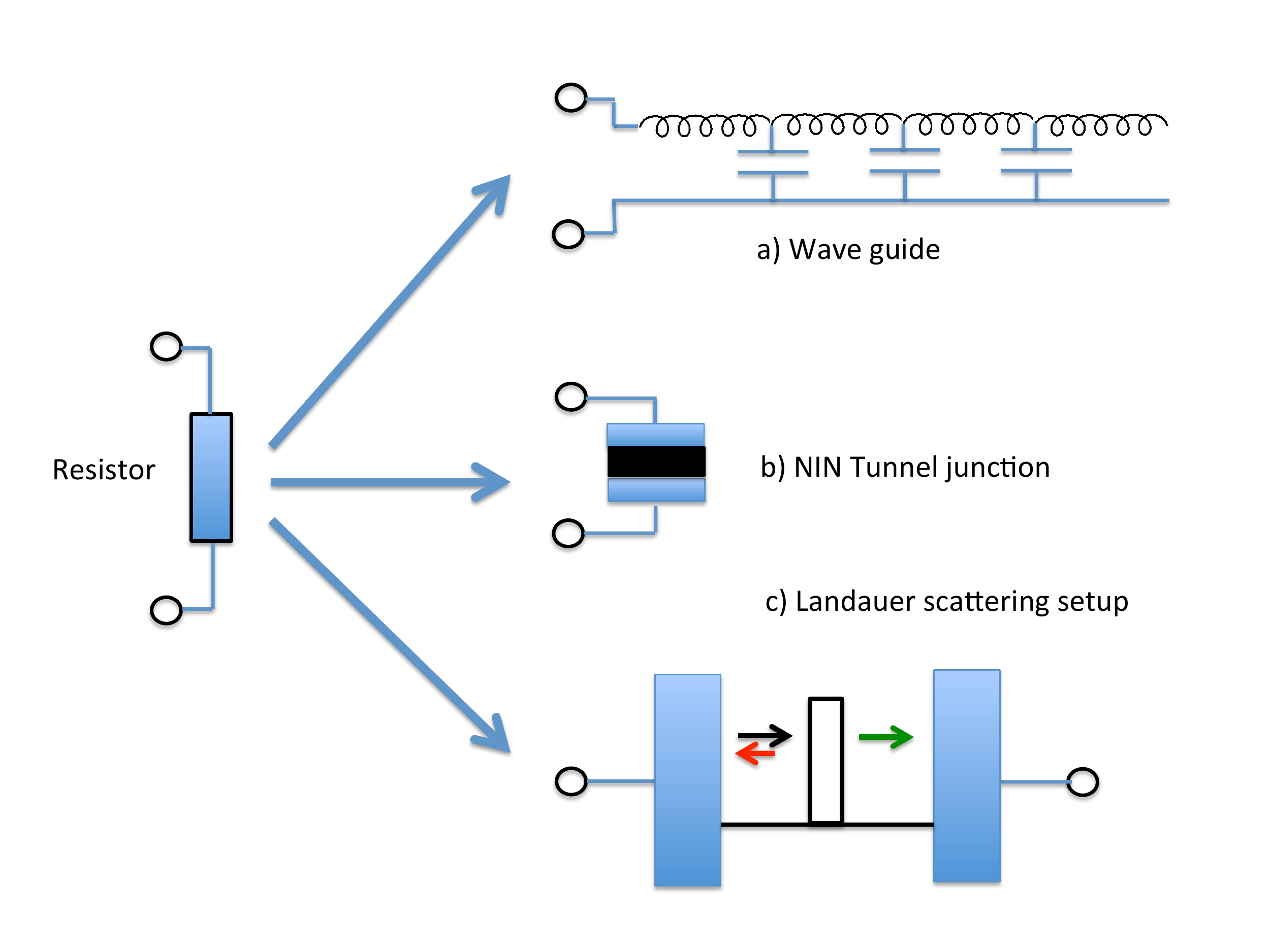}
\caption{Three approaches to envisioning dissipation: a) an LC circuit (wave-guide), extracting energy from the system; b) a dissipative tunnel junction, suitable for the AES picture; c) the Landauer picture consisting of a tunnel barrier (or a tunnel juction) 
coupled to dissipative reservoirs.}
\label{fig:Resistors}
\end{figure}

A paradigmatic framework to present a dissipative environment, in a way that connects to our preformed classical intuition, is to model Ohmic resistor quantum mechanically. We mention here three approaches:

1. The Caldeira-Leggett (CL) modelling~\cite{CaldeiraLeggett81}: One introduces an effective circuit consisting of an L-C transmission line 
(with infinitesimal imaginary term), that may extract energy and current from the bare quantum system.  (cf. Fig.~\ref{fig:Resistors}a)

2. The Ambegaokar-Eckern-Sch\"on (AES) modeling~\cite{AES_PRL,AES_PRB}: Here we model a tunnel junction 
(cf. Fig.~\ref{fig:Resistors}b) , assuming explicitly that  its transparency is low, hence only lowest order contributions in the tunneling should be accounted for. The resulting Hamiltonian represents reservoir degrees of freedom that give rise to dissipation. 
Traditional applications of the CL picture employed extended coordinates (this, however, is not a must; the CL action in the case of a spin degree-of-freedom consists of compact coordinates). By contrast the AES approach introduces compact (periodic) coordinates.  

3. The Landauer picture~\cite{Landauer70,Buettiker86,ImryBook}. Here one models the resistor by a tunnel barrier (of arbitrary transparency) (cf. Fig.~\ref{fig:Resistors}c for the single channel case). The contribution of this tunnel barrier to the resistance is given by $R/(1-R)$, where the reflection probability off the barrier is equal to the modulus square of the reflection  amplitude, $R=|r|^2$. This elastic backscattering process yields  the magnitude of the resistor; the actual inelastic dissipation takes place in the connected reservoirs. Such a model has been discussed, for example, in Ref.~\cite{NazarovBlanterBook}. We shall not consider this picture here.

The outline of this paper is the following: in Section~\ref{sec:CLvsAES} we will briefly review earlier works, emphasizing  the difference between the CL and AES approaches to  dissipative dynamics, focusing on charge dynamics. The gauge symmetry underlying  charge transport is U(1). In Section~\ref{sec:QDStoner} we will recall the physics of a quantum dot (QD) tuned to be near (but below) the Stoner instability. 
As such, the QD  supports large magnetization. Ignoring fluctuations in the magnitude of the spin, the spin degree of freedom possesses an SU(2) symmetry. 
The coupling of such a QD to external leads  gives rise to dissipation, 
which is formulated and studied within the framework of the AES (Section~\ref{Sec:SU2AES}). 
In Section~\ref{Sec:SU2CL} we first compare our AES analysis for the spin case to our results obtained within the CL framework. 
We then note that this AES vs. CL contrast differs from the AES vs. CL in the standard charge U(1) case. 
We conclude in Section~\ref{Conclusions}.

\section{Caldeira-Leggett versus Ambegaokar-Eckern-Sch\"on: the Charge U(1) Case}
\label{sec:CLvsAES}

We consider the dynamics associated with current through a resistor, and compare the 
two paradigmatic representation thereof: CL and AES.

\subsection{CL action}
The CL action of a current biased linear resistor (modeled as a transmission line) reads
\begin{equation}
i{\cal S}_{CL} = - \int d t_1 d t_2\, \alpha(t_1,t_2)\frac{\left[\varphi(t_1)-\varphi(t_2)\right]^2}{2} + i {\cal S}^{source}\ .
\label{eq:CLaction}
\end{equation}
Here the dimensionless phase variable $\varphi(t)$ represents the effective flux variable 
$\Phi(t)$ via $\varphi = 2\pi (\Phi/\Phi_0)$, where $\Phi_0 = h/e$ is the flux quantum. The voltage 
across the resistor is given by $V = d\Phi/dt$, and $\Phi$ is the degree-of-freedom canonically 
conjugate to the charge that has flown through the resistor $Q = \int dt\, I$. 
In (\ref{eq:CLaction}) $\alpha(t_1,t_2)$ is the kernel of the Ohmic bath~\cite{CaldeiraLeggett81}. 
Dropping the time-local terms (important for avoiding renormalization of the non-dissipative part of the action) we obtain 
\begin{equation}
i{\cal S}_{CL} =  \int d t_1 d t_2\,  \alpha(t_1,t_2)\varphi(t_1)\varphi(t_2) + i {\cal S}^{source}\ .
\label{eq:CLaction12}
\end{equation}
Note that in Keldysh notation this action may be written as 
\begin{eqnarray}
&&i{\cal S}_{CL} = i {\cal S}^{source} \nonumber \\&&+ 2  \int\limits_{-\infty}^{\infty} d t_1 \int\limits_{-\infty}^{\infty}  d t_2\,\,
\left(\begin{array}{cc}
\varphi_c(t_1) & \frac{\varphi_q(t_1)}{2}\end{array}
\right)
\left(
\begin{array}{cc}
0 & \alpha^A\\
\alpha^R & \alpha^K
\end{array}
\right)_{(t_1-t_2)}
\left(\begin{array}{c}
\varphi_c(t_2) \\ \frac{\varphi_q(t_2)}{2}\end{array}
\right)\ ,\nonumber\\
\label{eq:CLactionKeldysh}
\end{eqnarray}
where $c,q$ refer to the classical and quantum components on the Keldysh 
contour~\cite{KamenevBook}~\footnote{Unlike Ref.~\cite{KamenevBook} we use 
the convention $\varphi_c = (\varphi_u + \varphi_d)/2$, $\varphi_q = \varphi_u - \varphi_d$, where $u,d$ refer to the 
forward and the backward parts of the Keldysh contour respectively.}. The subscripts $R,A,K$ refer to the retarded, advanced, and Keldysh components of the matrix. 

Employing the relation between the retarded and the advanced components of the kernel $\alpha$, 
$\alpha^A(t_2,t_1)=- \left[\alpha^R(t_1,t_2)\right]^*$ we may write the action as 
\begin{equation}
\label{eq:SCLRKSource}
i{\cal S}_{CL} = i{\cal S}_{CL}^R + i{\cal S}_{CL}^K + i{\cal S}^{source}
\end{equation}
with 
\begin{equation}
i{\cal S}_{CL}^R  = 2 i \int d t_1 d t_2 \left[{\rm Im}\,\alpha^R(t_1-t_2)\right]\varphi_q(t_1)\varphi_c(t_2)\ , 
\end{equation}
\begin{equation}
i{\cal S}_{CL}^K  = \frac{1}{2} \int d t_1 d t_2 \alpha^K(t_1-t_2)\varphi_q(t_1)\varphi_q(t_2)\ , 
\end{equation}
and
\begin{equation}
i{\cal S}^{source}  = i  \int d t  I_{ex}(t) \frac{\Phi_0}{2\pi}\,\varphi_q(t)\ , 
\end{equation}
One may~\cite{ASchmid82_Langevin} rewrite the Keldysh term of the action, employing the decoupling 
\begin{equation} 
e^{\frac{i}{\hbar}{\cal S}_{CL}^K} = 
\int {\cal D}\xi\, e^{\frac{i}{\hbar}\int dt \,\hbar \xi(t) \varphi_q(t)}\,e^{\frac{1}{2}\int dt_1 dt_2 \hbar\,\left[\alpha_K\right]^{-1}_{t_1,t_2}\xi(t_1)\xi(t_2)}\ .
\end{equation}
It follows that 
\begin{equation}
\langle \xi(t_1)\xi(t_2) \rangle = \frac{1}{\hbar} \,\alpha^K(t_1,t_2)\ .
\end{equation}
The resulting Langevin equation-of-motion is obtained by calculating the variation $i \delta {\cal S}_{CL}/\delta \varphi_q(t) = 0$. The equation obtained is  
\begin{equation}
\label{eq:CLLangevinEq}
\frac{\dot \Phi_c(t)}{R}=I_{ex} + \delta I(t)\ ,
\end{equation}
where $\delta I(t) \equiv e \xi(t)$ represents stochastic current noise. We note that the 
{\it noise is additive}, and is not affected by the bias current.

In deriving Eq.~\ref{eq:CLLangevinEq} we have used the fact that the dissipative bath 
has an Ohmic spectrum~\cite{CaldeiraLeggett81}, implying that 
\begin{equation}
\label{eq:alphaRtomega}
{\rm Im}\, \alpha^R(t) = \frac{1}{2}\,\frac{1}{R}\,\frac{\hbar^2}{e^2} \delta'(t)\quad {\rm or}\quad  
{\rm Re}\, \alpha^R(\omega) = \frac{1}{2}\,\frac{1}{R}\,\frac{\hbar^2}{e^2} \omega\ ,
\end{equation}
where $R$ is the resistance. The variation over the retarded part of the action leads to 
\begin{equation}
\label{alphaRtomega}
\frac{i \delta S^R_{CL}}{\delta \Phi_q(t_1)} = \frac{2\pi}{\Phi_0}\, 2 i \int d t_2 \left[{\rm Im}\,\alpha^R(t_1-t_2)\right]
\left[2\pi\frac{\Phi_c(t_2)}{\Phi_0}\right]=\frac{i}{R}\,\dot \Phi_c(t_1)\ .
\end{equation}
The Fourier transform of the current noise correlator is given by
\begin{equation}
\langle \delta I(t_1) \delta I(t_2)\rangle_\omega = \frac{e^2}{\hbar}\,\alpha^K(\omega) \ .
\end{equation}
At equilibrium 
\begin{equation}
\label{alphaKRA}
\alpha^K(\omega) = \left[\alpha^R(\omega)-\alpha^A(\omega)\right]\,\coth\frac{\hbar\omega}{2k_{\rm B} T}\ .
\end{equation}
The fluctuation-dissipation theorem follows from Eqs.~(\ref{alphaRtomega}) and (\ref{alphaKRA}).
\begin{equation}
\langle \delta I(t_1) \delta I(t_2)\rangle_\omega = \frac{\hbar \omega}{R} \,\coth\frac{\hbar\omega}{2k_{\rm B} T} \ .
\end{equation}
We note that the additivity of the noise and its independence of the bias current (Eq.~\ref{eq:CLLangevinEq}) imply that the noise 
is independent of $I_{ex}$, i.e., absence of shot noise.

\subsection{AES action}
The AES action now is given by 
\begin{equation}
\label{SAESgeneral}
i{\cal S}_{AES} =- \int d t_1 d t_2\, \alpha(t_1,t_2)\left(1-\cos\left[\varphi(t_1)-\varphi(t_2)\right]\right)+ i {\cal S}^{source}\ .
\end{equation}
The source term is the same as in the previous case. Similarly to the CL case, Eq.~(\ref{eq:SCLRKSource}), 
one may write the action as 
\begin{equation}
\label{eq:SAESRKSource}
i{\cal S}_{AES} = i{\cal S}_{AES}^R + i{\cal S}_{AES}^K + i{\cal S}^{source}\ .
\end{equation}
The retarded part is essentially identical to that in the CL case, having to do with the fact that 
$t_1$ and $t_2$ are very close to each other (cf. Eq.~(\ref{eq:alphaRtomega})), which allows us to expand the 
$\cos(\dots)$ term in Eq.~(\ref{SAESgeneral}). 
The Keldysh term, though, is very different:
\begin{eqnarray}
&&i{\cal S}_{AES}^K  = \frac{1}{2} \int d t_1 d t_2 \alpha^K(t_1-t_2)\nonumber\\
&&\left\{\left[\cos\varphi(t_1)\right]_q\left[\cos\varphi(t_2)\right]_q +
\left[\sin\varphi(t_1)\right]_q\left[\sin\varphi(t_2)\right]_q
\right\}\ , 
\end{eqnarray}
Decoupling the action, employing two auxiliary fields, $\xi_1$ and $\xi_2$, one obtains~\cite{AES_PRB}
\begin{eqnarray} 
e^{\frac{i}{\hbar}{\cal S}_{AES}^K} = 
\int {\cal D}\xi_1{\cal D}\xi_2\, && e^{\frac{i}{\hbar}\int dt \,\hbar \left( \xi_1(t) \left[\cos\varphi(t)\right]_q 
+\xi_2(t) \left[\sin\varphi(t)\right]_q\right)}\nonumber\\ 
\times &&e^{\frac{1}{2}\int dt_1 dt_2 \hbar\,\left[\alpha^K\right]^{-1}_{t_1,t_2}\left(\xi_1(t_1)\xi_1(t_2)+\xi_2(t_1)\xi_2(t_2)\right)}\ .
\end{eqnarray}
The resulting equation-of-motion for the AES action is
\begin{equation}
\label{eq:AESLangevinEq}
\frac{\dot \Phi_c(t)}{R}=I_{ex} - e \xi_1 \sin\left(2\pi\frac{\Phi_c}{\Phi_0}\right) + e \xi_2 \cos\left(2\pi\frac{\Phi_c}{\Phi_0}\right)\ .
\end{equation}
This equation can be cast into the form of Eq.~(\ref{eq:CLLangevinEq}) by writing
$\delta I(t) = \delta I_1(t) + \delta I_2(t)$ with 
the two independent terms of current fluctuations defined as
\begin{equation}
\label{eq:deltaI12}
\delta I_1 = - e \xi_1 \sin\left(2\pi\frac{\Phi_c}{\Phi_0}\right)\quad,\quad \delta I_2 = e \xi_2 \cos\left(2\pi\frac{\Phi_c}{\Phi_0}\right)\ .
\end{equation}

The equation-of-motion (\ref{eq:AESLangevinEq}) implies that the noise is non-additive, as can be shown explicitly from the following iterative procedure. The zeroth iteration 
gives $\Phi_c = V t$, where $V= I_{ex}R$. Next we introduce a correction $\Phi_c = Vt + \delta \Phi_c$ and obtain 
\begin{equation}
\label{eq:AESLangevinEq1Iteration}
\frac{\delta \dot \Phi_c(t)}{R}=-e \xi_1 \sin\left(2\pi\frac{Vt + \delta\Phi_c}{\Phi_0}\right) + e \xi_2 \cos\left(2\pi\frac{Vt +\delta \Phi_c}{\Phi_0}\right)\ .
\end{equation}
The first iteration consists in dropping $\delta \Phi_c$ in the r.h.s. of Eq.~(\ref{eq:AESLangevinEq1Iteration}).
The resulting stochastic terms give rise to shot noise~\cite{AES_PRB} (unlike the CL equation-of-motion). 
For $eV \gg k_{B}T$ we find 
\begin{equation}
\langle \delta I_1(t_1) \delta I_1(t_2)\rangle_{\omega\rightarrow 0} = \langle \delta I_2(t_1) \delta I_2(t_2)\rangle_{\omega\rightarrow 0}  = \frac{1}{2}\, e\,\frac{V}{R}\ .
\end{equation}

\section{A Quantum Dot  near the Stoner Phase Transition}
\label{sec:QDStoner}
Over the past few decades the physics of quantum dots has become a focal point of research in nanoelectronics. The introduction of the ”Universal Hamiltonian“
\cite{PhysRevB.62.14886,AleinerBrouwerGlazman,alhassidreview,KamenevAndreev2}
has made it possible to take into account the effects of
electron-electron (e-e) interaction within a QD in a controlled way. This approach is applicable for a normal-metal QD when 
the Thouless energy $E_{Th}$ and the mean single particle level spacing $\delta$ satisfy $g_{\rm QD} \equiv E_{Th}/\delta>>1$. Here 
$g_{\rm QD}$ is the dimensionless conductance of the QD. The single particle level spacing is given by $\delta\sim 1/(V \nu_0)$, where $V$ is the volume 
of the QD and $\nu_0$ is
its density of states (DoS) and therefore $\delta \sim 1/L^d$ for a d-dimensional QD. The Thouless energy, $E_{Th}$, is the inverse 
time-of-flight (or diffusion time) of an electron across the quantum dot.

Within this scheme interactions are split into a sum of three spatially independent contributions in the charging, spin-exchange, and Cooper channels. Ignoring the latter (see below) the charging term leads to the phenomenon of Coulomb blockade, while the spin-exchange term can drive the system towards the Stoner instability~\cite{Stoner}. In bulk systems the exchange interaction competes with the kinetic energy leading to Stoner instability. In finite size systems mesoscopic Stoner regime may be a precursor of bulk thermodynamic Stoner instability~\cite{PhysRevB.62.14886,AleinerBrouwerGlazman}: a new phase, intermediate between paramagnetic and ferromagnetic emerges, in which the total spin of the QD  is finite but not extensive (i.e., not proportional to the volume of the dot). The mesoscopic Stoner regime can be realized in QDs made of materials close to the thermodynamic Stoner instability. 

A quantum dot in the metallic regime, $g_{\rm QD}\gg 1$, is described by the universal Hamiltonian~\cite{PhysRevB.62.14886}:
\begin{equation}
H =H_0 + H_C+H_J+H_{\lambda} . \label{EqUnivHam_1.1}
\end{equation}
The noninteracting part of the universal Hamiltonian reads
\begin{equation}
H_0= \sum\limits_{\alpha,\sigma} \epsilon_{\alpha} a^\dag_{\alpha,\sigma} a^{\phantom \dag}_{\alpha,\sigma}\ ,\label{EqUnivHam_1.2}
\end{equation}
where $\epsilon_{\alpha}$ denotes the energy of a spin-degenerate (index $\sigma$) single 
particle level $\alpha$. The charging interaction term
\begin{equation}
H_C=E_C \left ( \hat{N} - N_0\right )^2
\end{equation}
accounts for the Coulomb blockade. Here, $E_C\equiv e^2/(2C)$ denotes the charging energy of the quantum dot with the self-capacitance $C$, $N_0$ represents the background charge, and
$\hat N =\sum_{\alpha,\sigma} a^\dag_{\alpha,\sigma} a^{\phantom \dag}_{\alpha,\sigma}$ is the operator of the total number of electrons of the dot. For the isolated quantum dot the total number of electrons is fixed and, therefore, the charging interaction term can be omitted. The term
\begin{equation}
H_J = -J {\bf \hat S}^2
\end{equation}
represents the ferromagnetic ($J>0$) exchange interaction within the dot where ${\bf \hat S}=
\sum_{\alpha} a^\dag_{\alpha,\sigma_{1}}{\bf S}_{\sigma_{1}\sigma_{2}}a^{\phantom \dag}_{\alpha,\sigma_{2}}$ is 
the operator of the total spin of the dot. Here ${\bf S}_{\sigma_{1}\sigma_{2}}
\equiv(1/2) \vec \sigma_{\sigma_{1}\sigma_{2}}$, where $\vec{\sigma}=(\sigma_{x}, \sigma_{y}, 
\sigma_{z})$ is a vector made of Pauli matrices. The interaction in the Cooper channel is described by 
\begin{equation}
H_\lambda = \lambda T^{\dag} T^{\phantom \dag}, \qquad T = \sum_{\alpha} a^{\phantom\dag}_{\alpha,\uparrow} a^{\phantom \dag}_{\alpha,\downarrow} .
\end{equation}
In what follows we do not take into account $H_\lambda$ for the following reasons. For the dots defined
in 2D electron gas the interaction in the Cooper channel is typically repulsive and, therefore, renormalizes 
to zero~\cite{AleinerBrouwerGlazman}. In the case of 3D quantum dots realized as small metallic grains, the interaction 
in the Cooper channel can be attractive, giving rise to interesting competition between superconductivity and 
ferromagnetism~\cite{Schechter,AlhassidSF1,AlhassidSF2}. In that case we assume that there is a weak magnetic field 
which suppresses the Cooper channel.

The starting point of our analysis of a dissipative Stoner QD (near the Stoner instability point) accounts for the QD Hamiltonian
\begin{equation}
H_{dot} = \sum_{\alpha,\sigma} \epsilon^{\phantom\dag}_\alpha  a^\dag_{\alpha,\sigma} a^{\phantom \dag}_{\alpha,\sigma} - 
J \bm{S}^2 \ ,
\end{equation}
In doing so we ignore possible correlations between the charging state and the spin configuration of the 
QD~\cite{boazgefen}.

We note that for isotropic spin exchange interaction (Heisenberg model) the mesoscopic Stoner phase extends over $1/2 \leq  J/\delta \leq 1$. For the anisotropic case~\cite{PhysRevLett.96.066805,PhysRevB.90.195308} the lower boundary of this inequality  slides towards 1, with no mesoscopic Stoner phase for Ising spin~\cite{boazgefen,PhysRevB.89.201304}.
For the isotropic case the ground state spin S is the integer value (for even number of electrons on the QD) or half-integer value (for odd number) that is closest to $J/2(\delta-J)$.  This value increases with increasing J and diverges for $J \rightarrow \delta$, which marks the onset of the macroscopic  Stoner ferromagnetic phase. 
Seemingly the problem is easy to tackle theoretically. The interaction terms of the universal Hamiltonian consist only of  zero mode (zero wave-number) contributions, which commute with each other. The inclusion of the exchange term renders the problem non-trivial though: the resulting action, which consists of Pauli matrices, is non-Abelian (more specifically, it is underlined by an SU(2) symmetry). Attempts  to study the problem from different  points of view included the Ising limit~\cite{boazgefen}, perturbation theory in the Ising 
anisotropy~\cite{PhysRevLett.96.066805}. An exact solution that employs states classified by the total number of electrons and the total 
spin~\cite{AlhassidRupp,PhysRevB.69.115331,TureciAlhassid} requires the calculation of Clebsch-Gordan coefficients which
is not an easy task. In this way Alhassid and Rupp have
found an exact solution for the partition function in the
absence of Zeeman splitting. Elements of their analysis
were then incorporated into a master equation analysis of
electric~\cite{AlhassidRupp,PhysRevB.69.115331}
and thermal~\cite{PhysRevB.81.205303} properties. 
Independently, a study of electron
transport through a QD for 
low temperatures ($T \ll \delta$)
was made in reference~\cite{UsajBaranger}. 
That analysis, accounting for the charging and exchange interactions, employed 
a master equation approach as well. 

An exact solution based on the 
Wei-Norman-Kolokolov approach had been presented in reference~\cite{JETPLetters.92.179},
and was then extended to include randomness-induced spectral fluctuations~\cite{PhysRevB.85.155311}.
The tunneling density of states and the spin susceptibility were calculated; other thermodynamic and linear response correlations 
are calculable as well. 
The study of shot noise near the Stoner point was reported in~\cite{SothmannKoenigGefen}.

We note that the exact solution approaches mentioned above, while elegant and powerful, are very difficult to generalize to more complex setups, in particular, to setups where external leads are added -- a common mean
for the introduction of dissipation.
An efficient approximation, which can be generalized to such setups, employs adiabatic approximation of the spin stochastic dynamics~\cite{SahaAnnals}.

\section{AES approach for SU(2) spin}
\label{Sec:SU2AES}

Our approach~\cite{PhysRevLett.114.176806} can be viewed as a generalization of the Landau-Lifschitz-Gilbert (LLG)-Langevin 
equation~\cite{Gilbert2004,BrownLLGLangevin}, central to the field of spintronics~\cite{RevModPhys.77.1375}, to 
a regime where quantum dynamics dominates. 
Stochastic LLG equations have been derived in numerous publications for both a localized spin 
in an electronic environment (a situation of the Caldeira-Leggett type)~\cite{PhysRevB.73.212501,PhysRevB.85.115440} and for a magnetization formed by itinerant electrons~\cite{ChudnovskiyPRL,BaskoVavilovPRB2009}.
In all these works the precession frequency was assumed to be lower than the temperature or the voltage, thus 
justifying the semi-classical treatment of the problem. In this regime the geometric phase
did not influence the Langevin terms. 

Our derivation here is technically close to that of Ref.~\cite{ChudnovskiyPRL}. However, in contrast 
to Ref.~\cite{ChudnovskiyPRL}, we do not limit ourselves to small deviations of the spin from the instantaneous 
direction, but rather consider the action on global trajectories covering the entire Bloch sphere.
 
To demonstrate the emergence of an AES-like effective action we consider 
a quantum dot with strong exchange interaction coupled to a normal lead. 
The Hamiltonian reads
$H = H_{dot} + H_{lead} + H_{tun}$.
The quantum dot is described by the magnetic part~\footnote{
Here we disregard the charging part of the ''universal'' Hamiltonian, having in mind, e.g., systems of the type considered in Refs.~\cite{ChudnovskiyPRL,BaskoVavilovPRB2009}. Consequently no Kondo physics is expected.} 
of the universal Hamiltonian~\cite{PhysRevB.62.14886} 
\begin{equation}\label{eq:Hdot}
H_{dot} = \sum_{\alpha,\sigma} \epsilon^{\phantom\dag}_\alpha  a^\dag_{\alpha,\sigma} a^{\phantom \dag}_{\alpha,\sigma} - 
J \bm{S}^2 + \bm{B}\bm{S}\ ,
\end{equation}
where 
$
\bm{S}\equiv (1/2) \sum_{\alpha,\sigma_1,\sigma_2}\,a^\dag_{\alpha,\sigma_1} \bm{\sigma}_{\sigma_1,\sigma_2}\, a^{\phantom \dag}_{\alpha,\sigma_2}
$
is the operator of the total spin on the quantum dot, $\bm{B}$ is the external magnetic field,  and $J>0$ is the corresponding ``zero mode'' ferromagnetic exchange constant. The Hamiltonian of the lead and that 
describing the tunneling between the dot and the lead are standard:
$H_{lead} =  \sum_{\gamma,\sigma} \epsilon^{\phantom\dag}_{\gamma}  c^\dag_{\gamma,\sigma} c^{\phantom \dag}_{\gamma,\sigma}$ and $H_{tun} =  \sum_{\alpha,\gamma,\sigma} V^{\phantom \dag}_{\alpha,\gamma} a^{\dag}_{\alpha,\sigma} c^{\phantom  \dag}_{\gamma,\sigma}  + h.c.$. We assume here a non-magnetic lead. Here $\gamma$ is the orbital quantum number describing eigenmodes of the lead.

We consider the Keldysh generating functional
${\mathcal {Z}} = \int D\bar \Psi D\Psi \,\exp{[i\,{\cal S}_\Psi]}$,  
where the Keldysh action is given by 
${\cal S}_\Psi  =  \oint_K  d t\, (i {\bar\Psi} \partial_t \Psi - H )$ 
(plus the necessary source terms which are not explicitly written).
Here, for brevity, $\Psi$ denotes all fermionic fields and the time $t$ runs along the Keldysh contour. 
After standard Hubbard-Stratonovich manipulations~\cite{KamenevBook,JETPLetters.92.179,SahaAnnals} decoupling the interaction term 
$- J \bm{S}^2$ we obtain ${\mathcal {Z}} = \int D \bm{\mathcal{M}} \,\exp{[i\,{\cal S}_M]}$, and
the action for the bosonic vector $\bm{\mathcal{M}}(t)$ reads
\begin{equation}\label{SPhi}
i {\cal S}_{M} =  
\mathrm{tr\;ln} \left[\left(
\begin{array}{cc}
G_{dot}^{-1} &  - {\hat V} \\
-{\hat V}^\dag & G_{lead}^{-1} 
\end{array}
\right)\right] - i\, \oint\limits_K d t\, \frac{|\bm{\mathcal{M}}|^2}{4J}\ .
\end{equation}
Here 
$G_{dot}^{-1}  \equiv [ i\partial_t - \epsilon_\alpha  -
(\bm{\mathcal{M}}(t)+\bm{B})\cdot\bm{\sigma}/2]$, while 
$G_{lead}^{-1}  \equiv  i\partial_t - \epsilon_{\gamma}$. 
Both $G_{dot}^{-1}$ and $G_{lead}^{-1}$ are matrices with time, spin, and orbital indexes. 
We introduce $\bm{M}(t) \equiv \bm{\mathcal{M}}(t) + \bm{B}$.
Expanding (\ref{SPhi}) in powers of the tunneling matrix $\hat V$ and re-summing we easily obtain
\begin{equation}
i {\cal S}_{M} =  \mathrm{tr\;ln} \left[G_{lead}^{-1} \right] +  \mathrm{tr\;ln} \left[G_{dot}^{-1} - \Sigma\right] 
- i\, \oint\limits_K d t\, \frac{|\bm{M}-\bm{B}|^2}{4J}\ ,
\end{equation}
where the self energy reads
$\Sigma \equiv {\hat V} G_{lead} {\hat V}^\dag$. The first term is trivial, i.e., it would never contain the source fields. Thus, it will be dropped in what follows. 

{\it Rotating frame.}
We introduce a unit length vector 
\begin{equation}
\bm{n}(t) = (\sin\theta\cos\phi, \sin\theta\sin\phi,\cos\theta)
\end{equation}
through
$\bm{M}(t) = M(t)\bm{n}(t)$ and transform to a coordinate system in which $\bm{n}$ coincides with the $z$-axis 
$\bm{n}(t)\cdot\bm{\sigma} =  R(t) \sigma_z R^{\dag}(t)$. This condition 
identifies the unitary rotation matrix $R$ as an element of ${\rm SU(2)/U(1)}$. Indeed, if we employ the Euler angle representation
\begin{equation}
R = \exp{\left[-(i\phi/2) \sigma_z\right]} \exp{\left[-(i\theta/ 2) \sigma_y\right]} \exp{\left[-(i\psi/2) \sigma_z\right]}\ ,
\end{equation} 
then the angles $\phi(t)$ and $\theta(t)$ determine the direction of $\bm{n}(t)$, while $\psi(t)$ is arbitrary, i.e., 
the condition $\bm{n}(t)\cdot\bm{\sigma} =  R \sigma_z R^{\dag}$ is achieved with any value of $\psi(t)$. 
Thus, $\psi$ represents the gauge freedom of the problem. We introduce, first, a shifted gauge field $\chi(t) \equiv  
\phi(t) + \psi(t)$.
This way a periodic boundary condition, e.g., in the Matsubara 
representation $R(\tau) = R(\tau + \beta)$, is satisfied for $\chi(\tau+\beta)=\chi(\tau)+4\pi m$ (The fact that $m$ is integer is intimately related to the spin quantization~\cite{AbanovAbanov}). 
We can always assume trivial boundary conditions for $\chi$, i.e., $m=0$. 
We keep this representation of the rotation matrix $R$ also for the Keldysh technique. 

We perform a transition to the rotating frame and obtain
\begin{equation}
i{\cal S}_{M} =  \mathrm{tr\;ln} \left[R^{\dag}\left(G_{dot}^{-1} - \Sigma \right)R\right] 
-i\, 
\oint\limits_K d t\, \left[\frac{M^2}{4J} -\frac{\bm{B}\bm{M}}{2J}\right]
\end{equation} 
(we omit the constant term $\propto |\bm{B}|^2$).
For the Green's function of the dot this gives 
$
R^{\dag}G_{dot}^{-1} R =  i\partial_t - \epsilon_\alpha  - 
M(t)\,\sigma_z/2 - Q$, 
where we define the gauge (Berry) term as
$
Q \equiv R^{\dag}(-i \partial_t ) R 
= Q_\parallel + Q_\perp$. 
Here
$
Q_\parallel \equiv 
[\dot \phi(1-\cos\theta) - \dot\chi]\,\sigma_z/2 
$
and
$
Q_\perp \equiv 
-
\exp{\left[i\chi\sigma_z\right]}
\left[\dot\theta\,\sigma_y - \dot \phi \sin\theta \,\sigma_x  \right]\,\exp{\left[i\phi\sigma_z\right]}/2
$. 
Note, that $Q$ depends on the choice of the gauge field $\chi$. Finally, we obtain
\begin{eqnarray}
i{\cal S}_{M} &=&\mathrm{tr\;ln} \left[G_{dot,z}^{-1} -Q - R^{\dag}\Sigma R \right]
-i\, \oint\limits_K d t \left[\frac{M^2}{4J} -\frac{\bm{B}\bm{M}}{2J}\right]\ ,
\end{eqnarray}
where $G_{dot,z}^{-1} \equiv  i\partial_t - \epsilon_\alpha - 
(1/2)\,M(t)\,\sigma_z$.

To find the semi-classical trajectories of the magnetization we need to consider paths 
$M(t)$, $\theta(t)$, $\phi(t)$ on the Keldysh contour such that the quantum components are small.  
The quantum ($q$) and classical ($c$) components of the fields are expressed in terms of the forward  ($u$) and backward ($d$) 
components~\cite{KamenevBook}, i.e.,  $\phi_q(t)=\phi_u(t)-\phi_d(t)$ and $\phi_c(t)=(\phi_u(t)+\phi_d(t))/2$.
Performing the standard Keldysh rotation~\cite{KamenevBook} we thus obtain 
\begin{eqnarray}\label{SPhiRotated}
i {\cal S}_{M} &=&\mathrm{tr\;ln} \left[\tilde G_{dot,z}^{-1}  - \tilde Q  - \tilde R^\dag\tilde \Sigma\tilde R \right] 
\nonumber\\
&+& i\,\int d t \,\frac{\bm{B} \bm{M}_q}{2J}- i\,\int d t \,\frac{M_cM_q}{2J}\ ,
\end{eqnarray}
where
$\tilde G_{dot,z}^{-1} \equiv \tau_x G_{dot,z}^{-1}$. 
The local in time matrix fields $Q(t)$ and $R(t)$ also acquire the $2\times 2$ matrix structure in the Keldysh space, 
e.g., $\tilde Q = Q_c \tau_x + Q_q \tau_0 /2$, where $\tau_{x,y,z,0}$ are the standard Pauli matrices.

{\it The adiabatic limit.}
Thus far we have made no approximations. The action (\ref{SPhiRotated}) governs both the dynamics of the magnetization 
amplitude $M(t)$ and of the magnetization direction $\bm{n}(t)$.
Here we focus on the case of a large amplitude $M$ (more precisely, $M$ fluctuates around a large average value $M_0$. Such a situation arises either on the ferromagnetic side of the Stoner transition or on the paramagnetic side, but very close to the transition. In the latter case, as was shown in Refs.~\cite{JETPLetters.92.179,SahaAnnals}, it is the 
integration out of the fast angular motion of $\bm{n}$ which creates an effective potential for $M$, forcing it to acquire a finite average value. More precisely the angular motion with frequencies $\omega  \gg \max{[T,B]}$ (we
adopt the units $\hbar=k_B=1$) can be integrated out, renormalizing the effective potential for the slow part of $M(t)$. The very interesting question of the dissipative dynamics of slow longitudinal fluctuations of $M(t)$ in the mesoscopic Stoner regime will be addressed elsewhere. 
Here we focus on the slow angular motion and substitute $M(t)=M_0$. Thus, the last term of (\ref{SPhiRotated}) can be dropped. We note that in the adiabatic limit we may neglect $\tilde Q_\perp$ as it contributes only in the second order in
$d\bm{n}/dt$~\cite{SahaAnnals}.

The idea now is to expand the action (\ref{SPhiRotated}) in both $\tilde Q$ (which is small due to the slowness of $\bm{n}(t)$) 
and $\tilde R^\dag\tilde \Sigma\tilde R$ (which is small due to the smallness of the tunneling amplitudes). 
A straightforward analysis reveals that a naive expansion 
to the lowest order in both violates the gauge invariance with respect to the choice of $\chi(t)$. 
One can show that the expansion in $\tilde R^\dag\tilde \Sigma\tilde R$ is gauge invariant only if all orders 
of $\tilde Q$ are taken into account, that is if $(\tilde G_{dot,z}^{-1}  - \tilde Q)^{-1}$ is used as zeroth order Green's function 
in the expansion. This problem necessitates a clever choice of gauge, such that 
$(\tilde G_{dot,z}^{-1}  - \tilde Q)^{-1}$ is as close as possible to $\tilde G_{dot,z}$, i.e., the effect of $\tilde Q$ is ``minimized''.

{\it Choice of gauge.}
As the action (\ref{SPhiRotated}) is gauge invariant we are allowed to choose the most convenient form of $\chi(t)$. 
We make the following choice
\begin{eqnarray}
\dot \chi_c(t) & = & \dot \phi_c(t)  \,(1-\cos\theta_c(t)) \ ,\nonumber \\
\chi_q(t) & = & \phi_q(t)\,(1-\cos\theta_c(t))\ ,
\label{KeldyshGaugeFix}
\end{eqnarray}
which satisfies the necessary boundary conditions, i.e., $\chi_q(t=\pm\infty)=0$.

Here we present a detailed justification of the gauge which is presented 
in Eq.~(\ref{KeldyshGaugeFix}).
Ideally we should have chosen a gauge 
that would lead to $Q_\parallel=0$. Seemingly, this might have been achieved with the choice  
$\dot \chi(t) = \dot \phi(t)  \,(1-\cos\theta(t))$ on both branches of the Keldysh contour. This choice, however, 
violates our desired boundary conditions as the integrals over $\dot \chi$ accumulated between $t=-\infty$ and $t=+\infty$ on the upper and on the lower Keldysh branches are different. Such a difference would show up as non-trivial 
boundary conditions on $\chi_q$ at either $t=-\infty$ or $t=+\infty$. In other words, had we 
selected $\dot \chi(t) = \dot \phi(t)  \,(1-\cos\theta(t))$ we should have violated the requirement $\chi_q(t=\pm \infty)=0$. We note, though, that to linear order in the quantum components the condition $\dot \chi(t) = \dot \phi(t)  \,(1-\cos\theta(t))$ yields 
$\dot \chi_q = \dot \phi_q (1-\cos\theta_c) + \theta_q \sin\theta_c\,\dot\phi_c$, leading to 
$\chi_q(t) = \int\limits^t \, dt' \left[\dot \phi_q(t') (1-\cos\theta_c(t')) + \theta_q(t') \sin\theta_c(t')\,\dot\phi_c(t')\right]
=\phi_q(t) (1-\cos\theta_c(t)) +  \int\limits^t \, dt' \,\sin\theta_c(t') \Big[\theta_q(t')\,\dot\phi_c(t') -\dot\theta_c(t')\,\phi_q(t')  \Big]$. The first term vanishes at $t=\pm \infty$ but not the last term. 
We thus include only the first term in $\chi_q$, leading to Eq.~(\ref{KeldyshGaugeFix}).
The gauge (\ref{KeldyshGaugeFix})
satisfies the boundary conditions and leads to the desired cancellation $Q_{\|}^{c}=0$, whereas the quantum 
component of $Q_\|$ remains nonzero:
\begin{equation}
\label{Qparq}
Q_{\parallel,q} =\frac{1}{2}\, \sigma_z\, \sin\theta_c \,\left[\dot \phi_c \theta_q  - \dot \theta_c \phi_q\right]\ .
\end{equation}
At the same time this choice allows for the expansion of the Keldysh action in the small $\phi_q$ and $\theta_q$ as 
there are no $\dot \phi_q$ terms remaining in (\ref{Qparq}).

{\it Berry phase (Wess Zumino Novikov Witten (WZNW) action).}
Expanding the zeroth order in $\tilde \Sigma$ term of the action (\ref{SPhiRotated}) to first order in $\tilde Q$ we obtain the 
well known in spin 
physics (see, e.g., Refs.~\cite{Volovik87,AbanovAbanov}) Berry phase (WZNW) action
\begin{equation}
i{\cal S}_{WZNW} =
-\frac{1}{2}\, \int dt\, \mathrm{tr} \left[ G^K_{dot,z}(t,t) Q_{\parallel,q}(t)\right]\ ,
\end{equation}
which after a straightforward calculation reads
\begin{eqnarray}
\label{WZNW}
i{\cal S}_{WZNW} =i S \,\int dt \,\sin\theta_c \,\left[\dot \phi_c \theta_q  - \dot \theta_c \phi_q\right]\ ,
\end{eqnarray}
where $S\equiv N(M_0)/2$ is the (dimensionless) spin of the dot. Here $N(M_0)$ is the number of orbital levels of the dot 
in the energy interval $M_0$ around the Fermi energy. Roughly $S = M_0 \bar \rho_{dot}/2$, where 
$\bar \rho_{dot}$ is the density of states averaged over the energy interval $M_0$. The effects of mesoscopic 
fluctuations of the density of states were considered in Ref.~\cite{PhysRevB.85.155311}.  

{\it AES action.} 
The central result of the current paper is the AES-like~\cite{AES_PRL,AES_PRB} effective action, 
which we obtain by expanding 
(\ref{SPhiRotated}) to the first order in $\tilde R^\dag\tilde \Sigma\tilde R$:
$i{\cal S}_{AES} =- \mathrm{tr} \left[\tilde G_{dot,z}
{\tilde R}^{\dag}\,\tilde \Sigma\,\tilde R  \right]$. 
This gives
\begin{eqnarray}\label{SAESMATRIX}
&&i{\cal S}_{AES} =- g \int dt_1 d t_2 \nonumber\\
&&\mathrm{tr} \left[\left(\begin{array}{cc}
R_c^{\dag}(t_1) & \frac{R_q^{\dag}(t_1)}{2}\end{array}
\right)
\left(
\begin{array}{cc}
0 & \alpha^A\\
\alpha^R & \alpha^K
\end{array}
\right)_{(t_1-t_2)}
\left(\begin{array}{c}
R_c(t_2) \\ \frac{R_q(t_2)}{2}\end{array}
\right)
 \right]\ ,
\end{eqnarray}
where $g \equiv (\hbar/e^2) (G_{\uparrow} + G_{\downarrow})/2$. Here 
$G_\sigma \equiv 2\pi\, (e^2/\hbar)\, |V|^2\,\rho_{lead}^{\phantom\sigma}\rho_{dot}^{\sigma}$
is the tunneling conduction of the spin projection $\sigma$,  
$\rho_{dot}^{\uparrow/\downarrow}$ are the densities of states at the respective $\uparrow$ and $\downarrow$
Fermi levels, whereas the 
density of states in the lead, $\rho_{lead}$, is spin independent.
The standard~\cite{AES_PRB} Ohmic kernel functions are given by
$\alpha^{R}(\omega)- \alpha^{A}(\omega)= 2 \omega$ and $\alpha^{K}(\omega) =2 \omega \coth(\omega/2T)$.
The action (\ref{SAESMATRIX})
strongly resembles the AES action~\cite{AES_PRB}, with $U(1)$ exponents $\exp{\left[i\varphi/2\right]}$ replaced by the $SU(2)$ 
matrices $R$. Fixing the gauge of $R$ is an essential part of our procedure.

{\it Semi-classical equations of motion.} From the effective action (\ref{SAESMATRIX}) we derive the 
semi-classical equation of motion. We follow the ideas proposed in Ref.~\cite{ASchmid82_Langevin}.
Using the representation 
$R = A_0 \sigma_0 + i A_x \sigma_x + i A_y \sigma_y + i A_z \sigma_z$, with 
$A_0 \equiv  \cos\left[\frac{\theta}{2}\right] \, \cos\left[\frac{\chi}{2}\right]$, 
$A_x \equiv \sin\left[\frac{\theta}{2}\right] \sin{\left[\phi-\frac{\chi}{2}\right]}$,
$A_y \equiv - \sin\left[\frac{\theta}{2}\right] \cos{\left[\phi-\frac{\chi}{2}\right]}$,
$A_z \equiv - \cos\left[\frac{\theta}{2}\right] \, \sin\left[\frac{\chi}{2}\right]$
we rewrite the AES action (Eq.~(\ref{SAESMATRIX})) as ${\cal S}_{AES} ={\cal S}_{AES}^R + {\cal S}_{AES}^K$, where
\begin{eqnarray}\label{SRAES}
i{\cal S}^R_{AES} &=& 
- 2 i g \int dt_1 d t_2 \, \left[{\rm Im}\,\alpha^R(t_1-t_2)\right]  \sum_{j} A_{j,q}(t_1) A_{j,c}(t_2)\ ,\nonumber\\
\end{eqnarray}
and
\begin{eqnarray}\label{SKAES}
i{\cal S}^K_{AES} &=& -  \frac{g}{2}
 \int dt_1 d t_2 \,\,\alpha_{K}(t_1-t_2)  \sum_{j} A_{j,q}(t_1) A_{j,q}(t_2)\ .\nonumber\\
\end{eqnarray}
Here $j=0,x,y,z$.
The Keldysh part of the action (\ref{SKAES}) leads to random Langevin forces. This can be shown~\cite{ASchmid82_Langevin} using 
the Hubbard-Stratonovich transformation 
\begin{eqnarray}
&&e^{i{\cal S}_{AES}^K} =\int \left(\prod_{j=0,x,y,z} D \xi_j\right)  \times\nonumber\\&& 
\exp\left[\int dt \left\{i\sum_{j=0,x,y,z} \xi_j A_{j,q}\right\} +i{\cal S}_\xi \right]\ ,
\end{eqnarray}
where the action ${\cal S}_\xi$ is given by
\begin{equation}
i{\cal S}_\xi = -\frac{1}{2g}\, \sum_j \int dt_1 dt_2  \left[\alpha^{K}\right]^{-1}_{(t_1-t_2)}  \xi_j(t_1)\xi_j(t_2)\ .
\end{equation}
In other words, 
$\langle \xi_j(t_1) \xi_k(t_2)\rangle =  \delta_{jk}\,g\,\alpha^{K}(t_1-t_2)$ and $\langle \xi_j \rangle =0$.
We obtain the Langevin equations Eq.~(\ref{LLGLangevin}) from $\delta {i\cal S}_{total}/\delta \phi_q(t)=\delta i{\cal S}_{total}/\delta \theta_q(t)=0$, where 
$i {\cal S}_{total} \equiv i{\cal S}_{B}+i{\cal S}_{WZNW} +i{\cal S}^R_{AES} +\int dt\, \sum_j i \xi_j A_{j,q}$. 
Here $
i{\cal S}_{B} =- i S\gamma\,B\int dt \, \sin\theta_c\,\theta_q$ is the action related to the magnetic field (in $z$-direction). 
Prior to performing the variation of the action, the field $\chi$ is replaced according to the gauge fixing choice (Eq.~(\ref{KeldyshGaugeFix})). Finally, we use 
$\alpha_{R}^{''}(t) = (\partial_t +C) \delta(t)$ (the constant $C$ is important for causality but drops in our calculation) and obtain the following equations of motion:
\begin{eqnarray}
\dot\theta_c +  \tilde g\,
\sin\theta_c \dot \phi_c = \eta_\theta\ ,\nonumber \\
\sin\theta_c \left(\dot \phi_c - \gamma B\right)  - \tilde g\, \dot\theta_c  = \eta_\phi \ .
\label{LLGLangevin}
\end{eqnarray}
Here $\tilde g \equiv \frac{g}{2S}$ and  $\gamma=(J\bar \rho_{dot})^{-1}$ is the ``gyro-magnetic'' constant of order unity. 
The Langevin forces (torques) are given by 
\begin{eqnarray}
\label{etas}
\eta_\theta =&\phantom{-}&\frac{1}{2S}\, \cos\frac{\theta_c}{2}\,\left[ \xi_x \,\cos\left(\phi_c-\frac{\chi_c}{2}\right) + \xi_y \,\sin\left(\phi_c-\frac{\chi_c}{2}\right)\right]
\nonumber\\
&-&\frac{1}{2S}\, \sin\frac{\theta_c}{2}\,\left[ \xi_z \,\cos\frac{\chi_c}{2} + \xi_0 \,\sin\frac{\chi_c}{2}\right]\ ,\nonumber\\
\eta_\phi  = &-& \frac{1}{2S}\, \cos\frac{\theta_c}{2} \left[\xi_x \,\sin\left(\phi_c-\frac{\chi_c}{2}\right) -
\xi_y \,\cos\left(\phi_c-\frac{\chi_c}{2}\right)\right] \nonumber\\
&-& \frac{1}{2S}\,\sin\frac{\theta_c}{2}\left[\xi_z\,\sin\frac{\chi_c}{2}-\xi_0\,\cos\frac{\chi_c}{2}\right]\ .
\end{eqnarray}
The l.h.s.~of Eqs.~(\ref{LLGLangevin}) represent the standard Landau-Lifshitz-Gilbert (LLG) equations~\cite{Gilbert2004} (without a random torque).  The r.h.s.~represent the random Langevin torque. The latter is expressed in terms of four independent stochastic variables $\xi_j$ ($j=0,x,y,z$), which satisfy  $\langle \xi_j(t_1) \xi_k(t_2)\rangle =  \delta_{jk}\,g\,\alpha^{K}(t_1-t_2)$ and $\langle \xi_j \rangle =0$. 
On the gaussian level, i.e., if fluctuations of $\theta_c$ and $\phi_c$ are neglected in Eqs.~(\ref{etas}), the Langevin forces $\eta_\theta$ 
and $\eta_\phi$ are independent of each other and have the same autocorrelation 
functions:
$\langle \eta_\theta(t_1)\eta_\phi(t_2)\rangle = 0$ and $\langle \eta_\theta(t_1)\eta_\theta(t_2)\rangle = 
\langle \eta_\phi(t_1)\eta_\phi(t_2)\rangle$.
We emphasize that, in general, the noise depends on the angles $\theta_c$ and $\phi_c$ leading to complicated 
dynamics within Eqs.~(\ref{LLGLangevin}). In the classical domain, i.e., for frequencies 
much lower than $T$, we can approximate $\langle \xi_j(t_1) \xi_k(t_2)\rangle = 4gT \delta(t_1 - t_2)\, \delta_{jk}$. 
Then $\langle \eta_\phi(t_1)\eta_\phi(t_2) \rangle =\langle \eta_\theta(t_1)\eta_\theta(t_2) \rangle= (gT/S^2) \delta(t_1 - t_2)$. 
Thus, the situation is simple and we reproduce Ref.~\cite{BrownLLGLangevin}. 

{\it Effective temperature.} In the quantum high-frequency domain the situation is different. We cannot interpret the four independent 
fields $\xi_n$ as representing the components of a fluctuating magnetic field.  
A close inspection of equations~(\ref{LLGLangevin}) shows that in the regime of weak dissipation, 
$S\gg 1$ and $\tilde g \ll 1$, 
the spin can precess with frequency $\tilde B\equiv \frac{\gamma B}{1+\tilde g^2}$ at an almost constant $\theta$ for a long time of order (shorter than) $(\tilde g \tilde B)^{-1}$. For such time scales we can approximate $\phi_c = \tilde B t$ and 
$\chi_c = (1-\cos\theta_c)\phi_c = (1-\cos\theta_c)\tilde B t$. Thus the Langevin fields $\xi_n$ in (\ref{etas}) are multiplied by 
fast oscillating cosines and sines with frequencies $\omega_{cos}\equiv\tilde B \cos^2(\theta_c/2)$ and $\omega_{sin}\equiv\tilde B \sin^2(\theta_c/2)$.
Thus~\footnote{Here we have dropped non-stationary terms depending
on $t_1+t_2$}
\begin{eqnarray}
\langle \eta_{\phi,\theta}(t_1) \eta_{\phi,\theta}(t_2)\rangle_{\omega=0}=
\frac{g}{4S^2} 
\Bigl [ \cos^2({\theta_c}/{2}) 
\, \alpha_K \left(\omega_{cos}\right)
+
\sin^2({\theta_c}/{2}) 
\, \alpha_K \left(\omega_{sin}\right)\Bigr ]\ .
\label{etaeta}
\end{eqnarray}
In the quantum regime $T \ll \tilde B$ these correlation functions differ substantially from the classical ones, 
$\langle \eta_\phi(t) \eta_\phi(t')\rangle_{\omega=0}=\langle \eta_\theta(t) \eta_\theta(t')\rangle_{\omega=0} = g T /S^2$.
Thus, if the spin could be held for a long time on a constant $\theta_c=\theta_0$ trajectory  (one possible 
way to do so was proposed in Ref.~\cite{PhysRevLett.114.176806}), the diffusion would be determined 
by the quantum noise at frequencies $\omega_c$ and $\omega_s$, which are governed by the geometric phase.
More precisely, the spread of $\theta_c$ and $\phi_c$ (in the rotating frame) will be given 
by $(\Delta\theta)^2 = \sin^2\theta_0\,(\Delta \phi)^2 = D t$, where 
\begin{equation}
D = (g/S^2) T_{eff}\ ,
\label{DCoeff}
\end{equation} 
and the effective temperature is calculated from (\ref{etaeta}) to be 
\begin{eqnarray}
T_{eff} &=& \frac{\tilde B}{2}\,\cos^4\left (\frac{\theta_0}{2}\right )\coth\left[\frac{\tilde B}{2T}\cos^2\left (\frac{\theta_0}{2}\right ) \right]
\nonumber\\&+&
\frac{\tilde B}{2}\,\sin^4\left (\frac{\theta_0}{2}\right )\coth\left[\frac{\tilde B}{2T}\sin^2\left (\frac{\theta_0}{2}\right ) \right]\ .
\label{Teff}
\end{eqnarray}
We emphasize once again that this semi-classical analysis is valid for a highly non-equilibrium situation is which the spin 
is driven and is kept artificially at a trajectory with $\theta_c = \theta_0\neq 0$.

{\it Semi-classical approximation.} We are now ready to discuss the physical meaning of the semi-classical approximation, i.e., the expansion of the 
action (\ref{SAESMATRIX}) up to the second order in $\theta_q$ and $\phi_q$. 
The non-expanded action is periodic 
in both $\theta_q$ and $\phi_q$. The periodicity in $\phi_q$ corresponds to the quantization of the $z$ spin component 
$S_z = S \cos\theta_c$. By expanding we restrict ourselves 
to the long time limit, in which $S_z$ has already ''jumped'' many times by $\Delta S_z=1$ in the course of spin diffusion. 
We neglect, thus, higher than the second cumulants of spin noise (see, e.g., Ref.~\cite{Altland2010} for 
similar discussion of charge noise). We obtain, however, a correct second cumulant with down-converted quantum noise 
(similar to shot noise in the charge sector). This is due to the ''multiplicative noise'' character of our Keldysh action (\ref{SAESMATRIX})
similar to the original AES case~\cite{AES_PRB} (see also~\cite{GutmanPRB2005}).

{\it Equilibrium dynamics near $\theta_c=0$.} In the absence of external driving at $T\ll \tilde B$, Eqs.~(\ref{LLGLangevin}) lead 
to fast relaxation of the spin towards the north pole of the Bloch sphere, i.e., $\theta_c=0$. Here we show that the effective temperature introduced above looses its meaning in this case. Near the north pole the spherical coordinates are not adequate and we rewrite the Langevin 
equations~(\ref{LLGLangevin}) in cartesian coordinates. Namely, we define $x = \sin\theta_c \cos\phi_c \approx \theta_c \cos\phi_c$ and $y \approx \theta_c \sin\phi_c$. The new Langevin equations for $x$ and $y$ (valid for $x,y \ll1 $) read 
\begin{eqnarray}
&&\dot x =  -\tilde B y - g \tilde B x + \frac{1}{2S(1+\tilde g^2)}\left(\xi_x - g \xi_y\right)\ ,\nonumber\\
&&\dot y = \tilde B x - g \tilde B y+ \frac{1}{2S(1+\tilde g^2)}\left(\xi_y + g \xi_x\right)\ .
\end{eqnarray}
A straightforward analysis of these linear equations leads to the stationary widths (standard deviations) of order $\Delta x = \Delta y \sim 1/\sqrt{S}$. Taking into account the standard relation 
$\langle \vec S^2\rangle = \langle S_x^2\rangle +  \langle S_y^2\rangle +  \langle S_z^2\rangle = S(S+1)$, 
we observe that in the pure state $S_z = S$ the following relation holds 
$\langle S_x^2\rangle +  \langle S_y^2\rangle = S^2(\Delta x^2+\Delta y^2) = S$. Thus, fluctuations of order 
$\Delta x = \Delta y \sim 1/\sqrt{S}$ are purely quantum (they would be of this order also for $\Delta S_z \sim 1$) 
and the semiclassical analysis is inapplicable in this case.

\section{CL vs. AES}
\label{Sec:SU2CL}

In this Section we compare the SU(2) AES model described in Section~\ref{Sec:SU2AES} 
with the straightforward generalisation of the Caldeira-Legget model for the spin SU(2) case. 
We further notice the similarity  between \{the difference between  AES and  CL  in the U(1) case\} and  \{the difference between  AES and  CL in the SU(2) case\}.

\subsection{CL in the spin SU(2) case}
The Caldeira-Leggett action arises from the interaction of the type $H_{int} = {\bf h}\cdot{\bf n}$. Here 
${\bf n}\equiv {\bf S}/S$ and the vector field $\bf{h}$ represents isotropic fluctuations of an effective magnetic field with 
the Keldysh correlation function $\langle T_K h_n(t_{1}) h_m(t_{2})\rangle = g \alpha(t_1,t_2)\,\delta_{n,m}$, where the times $t_1$ and $t_2$ are on the Keldysh contour. The filed ${\bf h}$ can, in reality, be due to, e.g., the Kondo coupling of the localized spin $\bf S$ to the electron-hole continuum. The coupling constant $g$ is chosen so that the equations of motion are exactly the same as in the AES case, where $g$ was proportional to the tunneling conductance. Assuming the fluctuations are Gaussian one obtains the following effective action 
\begin{equation}
{\cal S}_{CL} = \frac{g}{2}\,\oint_K d t_1 \oint_K d t_2\, \alpha(t_1,t_2) \,\left(1- {\bf n}(t_1){\bf n}(t_2)\right)\ .
\end{equation}
The Keldysh analysis similar to that presented above produces 
again equations~(\ref{LLGLangevin}), however the Langevin terms look different:
\begin{eqnarray}
\label{eq:CLetathetaphi}
\eta_\theta  &=& \frac{1}{2S}\left(-\xi_x \sin\phi +\xi_y \cos\phi\right) \ ,\nonumber\\
\eta_\phi &=& \frac{\sin\theta}{2S}\,\xi_z -  \frac{\cos\theta}{2S}\,\left(\xi_x \cos\phi +\xi_y \sin\phi\right) \ .
\end{eqnarray}
Only three random fields $\xi_n$ ($n=x,y,z$) are needed. Their fluctuations are given by $\langle \xi_n(t_1) \xi_m(t_2)\rangle =  \delta_{nm}\,g\,\alpha^{K}(t_1-t_2)$. Exactly these equations are derived in Ref.~\cite{BrownLLGLangevin} before 
making the high temperature approximation, which makes the $\cos\phi$ and $\sin\phi$ factors in the 
right hand side unimportant. In contrast to the AES case we obtain
\begin{eqnarray}
\label{etaetaCL}
&&\langle \eta_\theta(t_1) \eta_\theta(t_2)\rangle_{\omega=0}=
\frac{g}{4S^2} \, \alpha^K \left(\omega= \tilde B\right)\ ,
\nonumber \\
&&\langle \eta_\phi(t_1) \eta_\phi(t_2)\rangle_{\omega=0}=\frac{g}{4S^2}\, 
\left[\cos^2\theta_c
\, \alpha^K \left(\omega= \tilde B\right)+
\sin^2\theta_c
\, \alpha^K \left(\omega= 0\right)\right]\ .\nonumber\\
\end{eqnarray} 
We observe that the diffusion is not isotropic is this case. That is, in the $\theta$-direction the diffusion 
is characterized by  $D_\theta = (g/S^2) T_{\theta}$,
where $T_{\theta} = \frac{1}{2}\,\tilde B\,\coth\left[\frac{\tilde B}{2T}\right]$. For the $\phi$-direction we obtain
$D_\phi = (g/S^2) T_{\phi}$ with 
$T_{\phi} = \frac{1}{2}\,\cos^2\theta\, \tilde B\,\coth\left[\frac{\tilde B}{2T}\right]+ \sin^2\theta\, T=\cos^2\theta\, T_{\theta}+  \sin^2\theta\, T$. We observe that $T_\theta > T_\phi$. This anisotropy is most pronounced for $\theta =\pi/2$ and $T \ll \tilde B$. Once again we emphasise that the above mentioned diffusion takes place in a highly non-equilibrium case of a spin being artificially held on a trajectory with constant $\theta \neq 0$. 
At equilibrium, as in the AES case, the semi-classical analysis is not applicable.

\subsection{Comparisons: CL and AES for U(1) and SU(2) cases}

Here we compare the  similarities and differences between the CL and the AES pictures  in the U(1) charge case,  with those in the SU(2) spin case.
In the U(1) case the semi-classical equation of motion can be cast in the form of Eq.~(\ref{eq:CLLangevinEq})
both for the CL and the AES models. Yet, the Langevin term, i.e., the fluctuating current, $\delta I$, is entirely
different in the two models at low temperatures, $k_{\rm B}T \ll e V$. In the CL case $\delta I = e \xi$ is produced by one stochastic variable $\xi$, whose noise spectrum is Ohmic at equilibrium. In the AES case one needs two independent stochastic variables $\xi_1$ and $\xi_2$ (see Eq.~\ref{eq:deltaI12}). Both these variables have equilibrium Ohmic noise, yet, due to the multiplicative oscillating factors in Eq.~(\ref{eq:deltaI12}), the noise of 
$\delta I$ at zero frequency is determined by the noise of $\xi_{1,2}$ at frequency $V$. This leads to the appearance of shot noise. 

Analogously, in the SU(2) spin case, both CL and AES models lead to the semi-classical stochastic LLG
equations of the form (\ref{LLGLangevin}).  The two Langevin terms (spin torques) 
$\eta_\theta$ and $\eta_\phi$ are, however, different in the two models. In the CL case 
$\eta_\theta$ and $\eta_\phi$ can be expressed (see Eq.~\ref{eq:CLetathetaphi}) via three independent 
stochastic variables $\xi_x,\xi_y,\xi_z$ (all having Ohmic equilibrium noise spectra). In comparison, in the AES 
case one needs four independent stochastic variables $\xi_x,\xi_y,\xi_z,\xi_0$ with Ohmic equilibrium spectrum (see Eq.~\ref{etas}).

In both SU(2) CL and SU(2) AES models the noise is multiplicative. That is, in both 
Eq.~(\ref{eq:CLetathetaphi}) and Eq.~(\ref{etas}), the independent Langevin variables are multiplied 
by trigonometric functions of the Euler angles $\theta$ and $\phi$. 
Thus, in both models,
the frequency shifts  are similar to those  leading to the shot noise in the U(1) case. 
Yet, these frequency shifts are very different between the CL and the AES cases. 
We consider again the example of the spin being held artificially at the trajectory with $\theta = \theta_0\neq 0$ and precessing with frequency $\tilde B$. In the CL model the spectrum of $\xi_z$ is not shifted, whereas 
the spectra of $\xi_x$ and $\xi_y$ are shifted by $\tilde B$. By contrast, in the AES case, the 
spectra of $\xi_x$ and $\xi_y$ are shifted by $\omega_{cos}\equiv\tilde B \cos^2(\theta_0/2)$
and the spectra of $\xi_z$ and $\xi_0$ are shifted by $\omega_{sin}\equiv\tilde B \sin^2(\theta_0/2)$.
Both these factors are geometrical and are determined by the Berry phase of the spin's trajectory.

\section{Conclusions}
\label{Conclusions}

In this paper we review the Caldera-Leggett (CL) and the Ambegaokar-Eckern-Sch\"on (AES) approaches to 
dissipation. We first remind the reader about the well known physics of dissipative charge dynamics 
underscored by the U(1) symmetry. Then, we provide an analogous treatment for the dissipative SU(2) spin dynamics. In both cases the Keldysh technique allows deriving the quasi-classical Langevin equations of motion. 
Except in the CL U(1) case, the noise is multiplicative, which leads to the admixture of the high frequency (quantum) 
noise components to the low frequency dynamics. This gives rise to shot noise in the charge dynamics and well as to the novel phenomenon of geometric dephasing in dynamics of large spins.

\section{Acknowledgements}

This work has been supported by GIF, by the ISF, by the Minerva Foundation, and by SFB/TR12 of Deutsche Forschungsgemeinschaft. The analysis of the CL SU(2) case was performed with support from 
Russian Science Foundation (Grant No. 14-42-00044).

\bibliographystyle{unsrt}
\bibliography{paper}

\begin{thebibliography}{10}

\bibitem{Schwinger61}
Julian Schwinger.
\newblock Brownian motion of a quantum oscillator.
\newblock {\em Journal of Mathematical Physics}, 2(3):407--432, 1961.

\bibitem{Keldysh65}
L.V. Keldysh.
\newblock Diagram technique for nonequilibrium processes.
\newblock {\em JETP}, 20:1018, 1965.

\bibitem{KamenevAndreev}
Alex Kamenev and Anton Andreev.
\newblock Electron-electron interactions in disordered metals: Keldysh
  formalism.
\newblock {\em Phys. Rev. B}, 60:2218--2238, Jul 1999.

\bibitem{KamenevBook}
A.~Kamenev.
\newblock {\em Field Theory of Non-Equilibrium Systems}.
\newblock Cambridge University Press, Cambridge, 2011.

\bibitem{AltlandBook}
Alexander Altland and Ben~D. Simons.
\newblock {\em Condensed Matter Field Theory}.
\newblock Cambridge University Press, second edition, 2010.
\newblock Cambridge Books Online.

\bibitem{ASchmid82_Langevin}
Albert Schmid.
\newblock On a quasiclassical langevin equation.
\newblock {\em J. of Low Temp. Phys.}, 49(5-6):609--626, 1982.

\bibitem{GardinerZollerBook}
C.~Gardiner and P.~Zoller.
\newblock {\em Quantum Noise: A Handbook of Markovian and Non-Markovian Quantum
  Stochastic Methods with Applications to Quantum Optics}.
\newblock Springer Series in Synergetics. Springer, 2004.

\bibitem{CaldeiraLeggett81}
A.~O. Caldeira and A.~J. Leggett.
\newblock Influence of dissipation on quantum tunneling in macroscopic systems.
\newblock {\em Phys. Rev. Lett.}, 46:211--214, Jan 1981.

\bibitem{AES_PRL}
Vinay Ambegaokar, Ulrich Eckern, and Gerd Sch\"on.
\newblock Quantum dynamics of tunneling between superconductors.
\newblock {\em Phys. Rev. Lett.}, 48:1745--1748, Jun 1982.

\bibitem{AES_PRB}
Ulrich Eckern, Gerd Sch\"on, and Vinay Ambegaokar.
\newblock Quantum dynamics of a superconducting tunnel junction.
\newblock {\em Phys. Rev. B}, 30:6419--6431, Dec 1984.

\bibitem{Landauer70}
Rolf Landauer.
\newblock Electrical resistance of disordered one-dimensional lattices.
\newblock {\em Philosophical Magazine}, 21(172):863--867, 1970.

\bibitem{Buettiker86}
M.~B\"uttiker.
\newblock Four-terminal phase-coherent conductance.
\newblock {\em Phys. Rev. Lett.}, 57:1761--1764, Oct 1986.

\bibitem{ImryBook}
Yoseph Imry.
\newblock {\em Introduction to Mesoscopic Physics}.
\newblock Oxford University Press, second edition, 2008.

\bibitem{NazarovBlanterBook}
Yuli.~V Nazarov and Yaroslav~M. Blanter.
\newblock {\em Quantum Transport: Introduction to Nanoscience}.
\newblock Cambridge University Press, 2009.

\bibitem{Note1}
Unlike Ref.~\cite {KamenevBook} we use the convention $\varphi _c = (\varphi _u
  + \varphi _d)/2$, $\varphi _q = \varphi _u - \varphi _d$, where $u,d$ refer
  to the forward and the backward parts of the Keldysh contour respectively.

\bibitem{PhysRevB.62.14886}
I.~L. Kurland, I.~L. Aleiner, and B.~L. Altshuler.
\newblock Mesoscopic magnetization fluctuations for metallic grains close to
  the {S}toner instability.
\newblock {\em Phys. Rev. B}, 62(22):14886--14897, Dec 2000.

\bibitem{AleinerBrouwerGlazman}
I.~L. Aleiner, P.~W. Brouwer, and L.~I. Glazman.
\newblock Quantum effects in {C}oulomb blockade.
\newblock {\em Physics Reports}, 358:309--440, July 2002.

\bibitem{alhassidreview}
Y.~Alhassid.
\newblock The statistical theory of quantum dots.
\newblock {\em Rev. Mod. Phys}, 72(4):895, October 2000.

\bibitem{KamenevAndreev2}
A.~V. Andreev and A.~Kamenev.
\newblock Itinerant ferromagnetism in disordered metals: A mean-field theory.
\newblock {\em Phys. Rev. Lett.}, 81(15):3199, October 1998.

\bibitem{Stoner}
E.~C. Stoner.
\newblock Ferromagnetism.
\newblock {\em Rep. Prog. Phys.}, 11:43, 1947.

\bibitem{Schechter}
M.~Schechter.
\newblock Spin magnetization of small metallic grains.
\newblock {\em Phys. Rev. B}, 70:024521, 2004.

\bibitem{AlhassidSF1}
S.~Schmidt, Y.~Alhassid, and K.~van Houcke.
\newblock Effect of a {Z}eeman field on the transition from superconductivity
  to ferromagnetism in metallic grains.
\newblock {\em Europhys. Lett.}, 80:47004, 2007.

\bibitem{AlhassidSF2}
S.~Schmidt and Y.~Alhassid.
\newblock Mesoscopic competition of superconductivity and ferromagnetism:
  Conductance peak statistics for metallic grains.
\newblock {\em Phys. Rev. Lett.}, 101:207003, 2008.

\bibitem{boazgefen}
B.~Nissan-Cohen, Y.~Gefen, M.~N. Kiselev, and I.~V. Lerner.
\newblock The interplay of charge and spin in quantum dots: The {I}sing case.
\newblock {\em Phys. Rev. B}, 84:075307, August 2011.

\bibitem{PhysRevLett.96.066805}
M.~N. Kiselev and Yuval Gefen.
\newblock Interplay of spin and charge channels in zero-dimensional systems.
\newblock {\em Phys. Rev. Lett.}, 96:066805, Feb 2006.

\bibitem{PhysRevB.90.195308}
A.~U. Sharafutdinov, D.~S. Lyubshin, and I.~S. Burmistrov.
\newblock Spin fluctuations in quantum dots.
\newblock {\em Phys. Rev. B}, 90:195308, Nov 2014.

\bibitem{PhysRevB.89.201304}
D.~S. Lyubshin, A.~U. Sharafutdinov, and I.~S. Burmistrov.
\newblock Statistics of spin fluctuations in quantum dots with {I}sing
  exchange.
\newblock {\em Phys. Rev. B}, 89:201304, May 2014.

\bibitem{AlhassidRupp}
Y.~Alhassid and T.~Rupp.
\newblock Effects of spin and exchange interaction on the {C}oulomb-blockade
  peak statistics in quantum dots.
\newblock {\em Phys. Rev. Lett.}, 91(5):056801, August 2003.

\bibitem{PhysRevB.69.115331}
Y.~Alhassid, T.~Rupp, A.~Kaminski, and L.~I. Glazman.
\newblock Linear conductance in {C}oulomb-blockade quantum dots in the presence
  of interactions and spin.
\newblock {\em Phys. Rev. B}, 69:115331, Mar 2004.

\bibitem{TureciAlhassid}
Hakan~E. T\"ureci and Y.~Alhassid.
\newblock Spin-orbit interaction in quantum dots in the presence of exchange
  correlations: An approach based on a good-spin basis of the universal
  hamiltonian.
\newblock {\em Phys. Rev. B}, 74:165333, Oct 2006.

\bibitem{PhysRevB.81.205303}
Gabriel Billings, A.~Douglas Stone, and Y.~Alhassid.
\newblock Signatures of exchange correlations in the thermopower of quantum
  dots.
\newblock {\em Phys. Rev. B}, 81:205303, May 2010.

\bibitem{UsajBaranger}
Gonzalo Usaj and Harold~U. Baranger.
\newblock Exchange and the {C}oulomb blockade:\quad{}peak height statistics in
  quantum dots.
\newblock {\em Phys. Rev. B}, 67:121308, Mar 2003.

\bibitem{JETPLetters.92.179}
I.S. Burmistrov, Yu. Gefen, and M.N. Kiselev.
\newblock Spin and charge correlations in quantum dots: An exact solution.
\newblock {\em JETP Letters}, 92(3):179--184, 2010.

\bibitem{PhysRevB.85.155311}
I.~S. Burmistrov, Yuval Gefen, and M.~N. Kiselev.
\newblock Exact solution for spin and charge correlations in quantum dots:
  Effect of level fluctuations and zeeman splitting.
\newblock {\em Phys. Rev. B}, 85:155311, Apr 2012.

\bibitem{SothmannKoenigGefen}
Bj\"orn Sothmann, J\"urgen K\"onig, and Yuval Gefen.
\newblock Mesoscopic {S}toner instability in metallic nanoparticles revealed by
  shot noise.
\newblock {\em Phys. Rev. Lett.}, 108:166603, Apr 2012.

\bibitem{SahaAnnals}
Arijit Saha, Yuval Gefen, Igor Burmistrov, Alexander Shnirman, and Alexander
  Altland.
\newblock A quantum dot close to {S}toner instability: The role of the {B}erry
  phase.
\newblock {\em Annals of Physics}, 327(10):2543, 2012.

\bibitem{PhysRevLett.114.176806}
Alexander Shnirman, Yuval Gefen, Arijit Saha, Igor~S. Burmistrov, Mikhail~N.
  Kiselev, and Alexander Altland.
\newblock Geometric quantum noise of spin.
\newblock {\em Phys. Rev. Lett.}, 114:176806, Apr 2015.

\bibitem{Gilbert2004}
T.~L. Gilbert.
\newblock A phenomenological theory of damping in ferromagnetic materials.
\newblock {\em IEEE Transactions on Magnetics}, 40(6):3443--3449, Nov 2004.

\bibitem{BrownLLGLangevin}
William~Fuller Brown.
\newblock Thermal fluctuations of a single-domain particle.
\newblock {\em Phys. Rev.}, 130:1677--1686, Jun 1963.

\bibitem{RevModPhys.77.1375}
Yaroslav Tserkovnyak, Arne Brataas, Gerrit E.~W. Bauer, and Bertrand~I.
  Halperin.
\newblock Nonlocal magnetization dynamics in ferromagnetic heterostructures.
\newblock {\em Rev. Mod. Phys.}, 77:1375--1421, Dec 2005.

\bibitem{PhysRevB.73.212501}
Hosho Katsura, Alexander~V. Balatsky, Zohar Nussinov, and Naoto Nagaosa.
\newblock Voltage dependence of {L}andau-{L}ifshitz-{G}ilbert damping of spin
  in a current-driven tunnel junction.
\newblock {\em Phys. Rev. B}, 73:212501, Jun 2006.

\bibitem{PhysRevB.85.115440}
Niels Bode, Liliana Arrachea, Gustavo~S. Lozano, Tamara~S. Nunner, and Felix
  von Oppen.
\newblock Current-induced switching in transport through anisotropic magnetic
  molecules.
\newblock {\em Phys. Rev. B}, 85:115440, Mar 2012.

\bibitem{ChudnovskiyPRL}
A.~L. Chudnovskiy, J.~Swiebodzinski, and A.~Kamenev.
\newblock Spin-torque shot noise in magnetic tunnel junctions.
\newblock {\em Phys. Rev. Lett.}, 101:066601, Aug 2008.

\bibitem{BaskoVavilovPRB2009}
Denis~M. Basko and Maxim~G. Vavilov.
\newblock Stochastic dynamics of magnetization in a ferromagnetic nanoparticle
  out of equilibrium.
\newblock {\em Phys. Rev. B}, 79:064418, Feb 2009.

\bibitem{Note2}
Here we disregard the charging part of the ''universal'' Hamiltonian, having in
  mind, e.g., systems of the type considered in Refs.~\cite
  {ChudnovskiyPRL,BaskoVavilovPRB2009}. Consequently no Kondo physics is
  expected.

\bibitem{AbanovAbanov}
A.~G. Abanov and Ar. Abanov.
\newblock Berry phase for a ferromagnet with fractional spin.
\newblock {\em Phys. Rev. B}, 65:184407, Apr 2002.

\bibitem{Volovik87}
G~E Volovik.
\newblock Linear momentum in ferromagnets.
\newblock {\em Journal of Physics C: Solid State Physics}, 20(7):L83, 1987.

\bibitem{Note3}
Here we have dropped non-stationary terms depending on $t_1+t_2$.

\bibitem{Altland2010}
Alexander Altland, Alessandro De~Martino, Reinhold Egger, and Boris Narozhny.
\newblock Transient fluctuation relations for time-dependent particle
  transport.
\newblock {\em Phys. Rev. B}, 82:115323, Sep 2010.

\bibitem{GutmanPRB2005}
D.~B. Gutman, A.~D. Mirlin, and Yuval Gefen.
\newblock Kinetic theory of fluctuations in conducting systems.
\newblock {\em Phys. Rev. B}, 71:085118, Feb 2005.

\end{thebibliography}

\end{document}